\input jnl.tex
\singlespace
\preprintno{UPR--0686-T}
\preprintno{UF-IFT-HEP--95-10}
\preprintno{NSF-ITP--95-72}
\preprintno{DOE/ER/40561-215--INT95--17-03}
\preprintno{April 1996}
\vskip 1.75cm
\centerline{\bf SOLAR CORE HOMOLOGY,}
\centerline{\bf SOLAR NEUTRINOS AND}
\centerline{\bf HELIOSEISMOLOGY$^{*}$}
\vskip 1.5cm
\centerline{\bf Sidney A. Bludman$^{\dagger}$}
\vskip .5cm
\centerline{\it Department of Physics, University of Pennsylvania}
\centerline{\it Philadelphia, Pennsylvania 19104 USA}
\vskip 1cm
\centerline{\bf Dallas C. Kennedy$^{\ddagger}$}
\vskip .5cm
\centerline{\it Department of Physics, University of Florida}
\centerline{\it Gainesville, Florida 32611 USA}
\vskip 1.75cm
\centerline{ABSTRACT}
\itemitem{}{\tenrm     
Precise numerical standard solar models (SSMs) now agree with one another and 
with helioseismological observations in the convective and outer radiative 
zones.  Nevertheless these models obscure how luminosity, neutrino production 
and $g$-mode core helioseismology depend on such inputs as opacity and nuclear 
cross sections.  
Although the Sun is not homologous, its inner core by itself is chemically
evolved and almost homologous,
because of its compactness, radiative energy transport, and $ppI$-dominated
luminosity production.  We apply luminosity-fixed homology transformations to 
the core to estimate theoretical uncertainties in the SSM and to obtain a broad 
class of non-SSMs, parametrized by central temperature and density and purely
radiative energy transport in the core.}
\vskip 1.25cm
\centerline{Published in {\bf the Astrophysical Journal}: 20 November 1996}
\footnote{}{$^*$~~An earlier version appeared as U. Pennsylvania preprint 
UPR--0615-T (1994)
and in {\it Proc. INT Solar Modeling Workshop} (March 1994).}
\footnote{}{$^\dagger$~~e-mail: bludman@bludman.hep.upenn.edu.}
\footnote{}{$^\ddagger$~~e-mail: kennedy@phys.ufl.edu.}

\vfill\eject

{\bf 1. Homology and the Solar Core}	
\bigskip
After more than three decades of nuclear cross section measurements, opacity
calculations, and detailed computer evolutionary calculations,
standard solar models (SSMs) with the same inputs 
now agree in their neutrino flux predictions to within about 1\%. The
theoretical models are now also consistent with precise $p$-mode 
helioseismological observations of the Sun's outer radiative zone $x \equiv 
r/R_\odot=0.26-0.71$ and convective zone $x > 0.71$.  If the $g$-mode
helioseismological oscillations have sufficient amplitude,
their observation
can be expected soon to calibrate the solar inner core, where 
the thermonuclear luminosity and neutrino production take place. While 
necessary in the complex convective 
zone and justified by the precise helioseismological
observations, the complexity and numerical form of precise SSMs obscure the
simplicity of the solar core and the determinants of solar neutrino fluxes. In 
order to understand standard and non-standard solar models, we return to the 
homology methods of Schwarzschild (1958) and Iben (1969 and 1991) used before 
the advent of fast computers, but with three new features.

Castellani {\it et al.} (1993) have found that changing input parameters by 
factors as large as two leads to only homologous changes over 60\% by mass of 
the Sun.
In this paper, we explain this remarkable homology and demonstrate that,  
while the entire Sun is certainly not homologous, the core is homologous enough 
to be parametrized by its central temperature, $T_c$ and density $\rho_c$. 
This $(T_c,\rho_c)$-parametrization subsumes all astrophysical effects of 
opacity, composition, and the $ppI$ nuclear cross section factors $S_{11}$ and 
$S_{33}$ (the $pp$ and ${}^3He-{}^3He$ reactions) into the two parameters 
$(T_c,\rho_c)$, one representation of the central boundary conditions of solar
structure.  Indeed, any standard or 
nonstandard solar model that depends principally on radiative energy transport 
can be parametrized by $(T_c,\rho_c)$ and the remaining nuclear cross section 
factors $S_{34}$ and $S_{17}$ (the ${}^3He-{}^4He$ and $p-{}^7Be$ 
reactions) (Degli'Innocenti 1994).
Even solar models with a non-standard low opacity or low metallicity $Z$ are all
essentially parametrized by $(T_c,\rho_c)$ or by $S_{11}$, the principal cross 
section factor determining $T_c$. (See Figure~2 of Hata {\it et al.} (1994)
or Figure~2 of Hata and Langacker (1995); see also Hata (1994).)  As the 
$\rho_c$ dependence of the neutrino fluxes
is weak (see section~4 below), Hata and Langacker (1994) were able to show that 
the 0.7\% theoretical uncertainty in $T_c$, 
together with the remaining nuclear cross section uncertainties, provide the 
same theoretical neutrino flux and rate uncertainties and correlations that 
Bahcall and Ulrich obtained from 1000 Monte Carlo SSM simulations.  (See 
Figures~2-4 and~6-8 of Hata and Langacker (1995).)  

The $(T_c,\rho_c)$ parametrization allows analytic estimation of the 
logarithmic derivatives $\beta_j(i) \equiv \partial \ln\phi(i)/\partial \ln S_j$
of the principal neutrino fluxes $\phi(i)$ with respect to input parameters 
$S_j$,  which Bahcall and Ulrich (1988) obtained from 1000 Monte Carlo SSMs 
calculated 
with {\it small} changes in input parameters.  Because homology makes these 
logarithmic derivatives constants, from any precise SSM, we can not only
estimate theoretical uncertainties, as did Bahcall and Ulrich, but we can now 
also extrapolate to non-SSMs, so long as the energy transport is primarily 
radiative.

Our application of homology differs from earlier ones in three ways: (1) We 
apply homology only to the solar inner core, not to the whole Sun; (2) we do 
not assume $\rho \sim T^3$ or any polytropic relation; and (3) we use homology 
to derive the dependence of {\it core} temperature and density on opacity,
nuclear energy generation, and mean molecular weight, at {\it fixed} luminosity,
instead of the dependence of {\it effective} temperature on luminosity, with 
fixed opacity and nuclear energy generation (Cox and Giuli 1968).

After accounting for the different energies released when $pp,$ ${}^7Be,$
${}^8B,$ and $CNO$ neutrinos are produced,
the known solar luminosity $L_\odot$ fixes the total photon energy production
and constrains the neutrino fluxes through the nuclear reactions:
$$\eqalign{
\phi(pp) + (0.967)\, \phi(Be) + (0.743)\, \phi(B) + 
(0.946)\, \phi(CNO)
= 6.48 \times 10^{10} \, {\rm cm}^{-2}{\rm s}^{-1}\quad ,}\eqno(1)
$$\noindent
for the four principal neutrino fluxes.  (See Appendix; Castellani {\it et al.}
(1993); Bludman {\it et al.} (1993); Hata and Langacker (1994).)  For the
SSM, the three $pp$ branches $I, II, III$ terminate in the ratio
83.7\%~:~16.3\%~:~0.02\%.  Differentiating this constraint and assuming these 
termination ratios continue to hold approximately, we have the constraint
$$\eqalign{
\beta_i(pp) + (0.079)\beta_i(Be) + (0.000071)\beta_i(B) + 
(0.0145)\beta_i(CNO) = 0\quad }\eqno(2)
$$\noindent
on the logarithmic derivatives of the principal neutrino fluxes with respect to
any input parameter $S_j$.  The logarithmic derivatives obtained by Bahcall
and Ulrich from their 1000 Monte Carlo SSMs satisfy this sum rule (Bahcall and
Ulrich 1988; Turck-Chi\` eze, 1988; Bludman {\it et
al.} 1993; Hata 1994; Hata {\it et al.} 1994; Hata and Langacker 1994, 1995).
For models that depart greatly from the SSM, the three termination ratios
in general change significantly from the values used here; in that case, the
new ratios can be used as the new constraint, for small deviations from the
new SM.  The numerical coefficients of equations~(1) and~(2) also change.

The central temperature and density are outputs characterizing solar models.
Along with the nuclear cross sections and chemical composition, given by the 
vector {\bf X} of element abundances by mass, they determine the neutrino 
fluxes.  
The logarithmic derivatives of these fluxes with respect to central temperature,
$\alpha(i) \equiv \partial \ln\phi(i)/\partial \ln T_c$, satisfy
$$\eqalign{
\alpha(pp) + (0.079)\alpha(Be) + (0.000071)\alpha(B) + 
(0.0145)\alpha(CNO) = 0\quad ,}\eqno(3)
$$\noindent
and are functionals of temperature, density, and composition, that can be 
approximated by power laws in temperature, if the weak $\rho_c$ dependence is
ignored.  If $\alpha(pp)=-1.2,$ then $\alpha(Be)=8\pm 2$, $\alpha(CNO)
= 34\pm 9,$ is consistent with this constraint.  Once the $ppII,$ $ppIII,$ and 
$CNO$ cycles are included, the luminosity constraint prevents the solar core 
from being strictly homologous and induces a small $\rho_c$ dependence.
\bigskip
{\bf 2. The Present Sun}
\bigskip
The luminosity- and neutrino-generating core and the outer radiative and 
convective zones are almost decoupled dynamically.  For this reason, the solar
model inputs of luminosity $L_\odot$ and radius $R_\odot$ almost 
separately determine the two outputs, the helium abundance $Y$ and the 
convective zone mixing length (Bahcall and Ulrich 1988; Turck-Chi\` eze {\it 
et al.} 1988).  This permits us to ignore the convective zone, 
for which accurate opacities and detailed numerical models are needed, and to
concentrate on the outer radiative zone and the inner core.  The radius 
$R_\odot$ being irrelevant to the inner core leaves only the luminosity 
$L_\odot$ and the mass $M_\odot$ as fundamental parameters.

The equation of hydrostatic equilibrium is
$$\eqalign{
-dP/\rho dr = g \equiv Gm/r^2}\eqno(4a)
$$\noindent
or
$$\eqalign{
\equiv 1/\lambda_P = g/(P/\rho)\quad ,}\eqno(4b)
$$\noindent
where $\lambda_P \equiv (-d\ln P/dr)^{-1}$ is the pressure scale height.  We 
define the stiffness (effective polytropic exponent) $\Gamma\equiv d\ln P/d\ln 
\rho \equiv 1+(1/n_{\rm eff})$ and effective polytropic index $n_{\rm eff} 
\equiv d\ln\rho/d\ln(P/\rho)$ and write the equation of state as $P/\rho = {\Re}
(T/\mu)(1+D)$, where ${\Re}$ is the gas constant, $\mu$ is the mean molecular 
weight and $D$ includes all corrections to the ideal gas equation of state.
Then $1 - \Gamma^{-1}\equiv d\ln(P/\rho)/d\ln P\equiv \nabla - \nabla_\mu$,
where the thermal gradient $\nabla\equiv d\ln T/d\ln P$ and chemical gradient
$\nabla_\mu\equiv
d\ln (\mu/(1+D))/d\ln P$.  If $n_{\rm eff}$ and $\mu/(1+D)$ were constant, the 
Sun would be a polytrope of index $n_{\rm eff}$ and thermal gradient $\nabla = 
1/(n_{\rm eff}+1)$. These conditions obtain in the convective and outer 
radiative zones of the Sun, but not in the chemically evolved and inhomogeneous
inner core.  The core structure is usually obtained by evolutionary models
that depend on the initial relative metallicity $Z/X$ for the proto-Sun and
the present age.

Figures~1 and 2 show the gradients $\nabla,~1-\Gamma^{-1}$ and $\Gamma$ as 
function of the dimensionless radius $x\equiv r/R_\odot$ and included mass 
$m/M_\odot$, derived from the SSM of Dearborn 1994.  (This model provides dense 
enough output to allow us to plot the $P,~\rho,~T$ logarithmic derivatives and
agrees well with other SSMs, such as the SSM of Bahcall and Pinsonneault 
(1992) with helium diffusion.)
The Sun's convective zone is an $n_{\rm eff}=3/2$ polytrope; the outer 
radiative zone,
approximately an $n_{\rm eff}=4.3$ polytrope.  The core, chemically evolved and
inhomogeneous, contains the Sun's luminosity production and the majority of
its mass.  This complex structure prevents the entire Sun from being 
homologous, even though luminosity production and opacity are approximately 
power laws in individual zones.

The simplicity of the passive outer radiative zone governs any matter-amplified
neutrino oscillations which may take place there.  The concentration of mass 
and luminosity production interior to this zone makes it 
approximately an $n_{\rm eff}=4.3$ polytrope, with density scale 
height $\lambda_\rho = \Gamma \lambda_P$ very nearly constant at $R_\odot/10.5$.
With this scale height, if neutrinos of energy $E$ and mass squared difference 
$\Delta m^2$ oscillate with vacuum mixing angle $\sin^2\theta$, the adiabaticity
is
$$\eqalign{
{\cal A} = 5.3 \times 10^7 (\Delta m^2(eV^2)/E(MeV))(\sin^2 2\theta/\cos 2\theta)\quad ,}\eqno(5)
$$\noindent
and the jump probability $P_j= \exp (-E_{NA}/E)$, with
$$\eqalign{
E_{NA} \equiv (\pi^2/2){\cal A} E= 3 \times 10^8 ~ \Delta m^2 (\sin^2 2\theta/\cos 2\theta)\quad .}\eqno(6)
$$\noindent
The constant density scale height of the $n=4.3$ polytropic outer radiative zone
is the principal property of the Sun effecting the Mikheyev-Smirnov-Wolfenstein
(MSW) neutrino oscillations.
The best small-angle MSW solution to the solar neutrino observations gives 
$E_{NA}\approx$ 13 MeV, so that $^8B$ neutrinos of energy E=10 MeV oscillate 
non-adiabatically with ${\cal A} = 0.3$ and $pp$ neutrinos of energy E=0.3 MeV 
oscillate adiabatically with ${\cal A}=10$.  

The solar inner core is reasonably inferred to have been initially convective
and therefore chemically homogeneous.  The thermal gradient was thus
initially adiabatic, $\nabla=\nabla_{\rm ad},$ decreasing since.
Meanwhile, the composition gradient $\nabla_{\mu}$ has been increasing 
from zero.
At the present epoch, these two evolutionary changes nearly compensate, $\nabla
\approx \nabla_\mu$, making $\Gamma\approx 1$, so that $T/\mu \approx$ constant 
(Figure~2) and the core is quite condensed.  Together with the thermal gradient 
$\nabla\approx \nabla_\mu \approx 1/3$, this makes $T^3/\rho$ constant to 
within 7\% for $x<0.3,~m/M_\odot<0.613$ (Bahcall and Ulrich 1988).  This 
accident of the 
present epoch makes $(\mu/(1+D))P\sim \rho^{4/3}$, which, in the chemically 
{\it inhomogeneous} core, is {\it not} an $n_{\rm eff}=3$ polytrope. 
Since $P/\rho\sim T/\mu\simeq$ constant in the core, the core structure is
nearly an ``isothermal'' $n_{\rm eff}=\infty$ polytrope with
$T/\mu$ rather than $T$ nearly constant.  We note for completeness that a
long-lived convective core (such as a present-epoch fully mixed core) would
not follow the simple homology of this paper, as the simple chemical 
stratification $T/\mu\simeq$ constant would be violated.

Although far from polytropic, the solar inner core is almost homologous, because
over the narrow range of density and temperature in 
the compact core, the nuclear energy generation and Rosseland mean opacity 
are approximated by the power laws,
$$\eqalign{
\varepsilon=\varepsilon_o\rho^{\lambda}T^{\nu},~~~~~~\kappa=\kappa_o\rho^n T^{-s}\quad ,}\eqno(7)
$$\noindent
where $\varepsilon_o, \kappa_o, \mu$ depend on the chemical composition.
Outside of the core ranges of density and temperature, these homologous forms
would still be valid, but with different exponents.  (Alternative forms of
energy transport, such as WIMPs, would destroy the homology altogether.)
The core, which
we define by radius $x < 0.26$ so as to contain 99\% of the energy generation 
and 50\% of the mass, has $T=(7.7-15.7)\times 10^6~K$, $\rho=(19-154)~{\rm g}
\ {\rm cm}^{-3},$ and a small pressure $P=P_c/12$ at its edge. In this range, 
we adopt 
$\lambda=1,~\nu\approx 4.24$ for the luminosity generation (dominantly but not
exclusively $ppI$ cycle) and fit the inner core OPAL opacity (Iglesias and 
Rogers 1991; Rogers 1992) by
$n\approx 0.43,~s\approx 2.47$.  Our inner core OPAL opacity fit is close to
the values $n=0.5,~s=2.5$ illustrated by Figure~10.4 of
B\" ohm-Vitense (1992).  For comparison, we also consider the Kramers 
opacity (Cox and Giuli 1968).
\bigskip
{\bf 3. Homology Applied to the Solar Core Only}
\bigskip
Homology is usually applied to zero-age main sequence (ZAMS), chemically 
homogeneous stars to derive power-law $L-M,~R-M,~L-T_{\rm eff}$ relations among 
luminosity, radius and {\it effective} temperature for {\it different} stars 
with the
{\it same} power-law opacity and luminosity generation.  As shown in Figure~3, 
taken from Kippenhahn and Weigert (1990), these homology relations are obeyed 
by models and actual stars on the ZAMS upper and lower main sequence, which 
respectively have convective cores or envelopes.  For these stars,
$L\sim M^{3.35}, ~R\sim M^{0.57}$ and $L\sim M^{3.2}, ~R\sim M^{0.8},$
respectively.

The Sun, however, is not chemically homogeneous, is mostly radiative, and is 
transitional between $pp$ and $CNO$ burning.  Even on the ZAMS, the Sun was not
homologous.  Figure~3 shows that, for $(0.3-1.3) M_\odot$ stars near the Sun, 
the exponents change rapidly with mass, with $L\sim M^{4.6}, ~R\sim M^{1.4}$
at the solar 
mass.  Therefore, we cannot apply homology to the entire present or ZAMS Sun.

On the other hand, because the core of the Sun has approximately power-law 
opacity and luminosity generation and has a low pressure boundary condition, 
$P\approx P_c/12$, we can apply homology to the isolated solar {\it core.}  
We do so in a different way: instead of deriving luminosity, mass and radius 
relations for a {\it family} of stars having the same opacity and luminosity 
generation, we derive the dependence of central temperature $T_c(\kappa,
\varepsilon,\mu)$ and density $\rho_c(\kappa,\varepsilon,\mu)$
on overall opacity, energy generation, and mean molecular weight,
for {\it one} star of {\it fixed} luminosity.  This enables us to scale 
from any given SSM to models of the same luminosity with different input 
parameters.

Assuming an ideal gas equation of state,
the mass conservation and hydrostatic equilibrium equations scale as
$$\eqalign{
\rho \sim m/r^3\quad ,\quad P \sim m^2/r^4\quad ,}\eqno(8)
$$\noindent
where $m$ the mass included inside radius $r$, so that
$$\eqalign{
m/r \sim P/\rho \sim T/\mu\quad ,\quad P_{\gamma}/P \sim T^4/P \sim (\mu^2m)^2
\quad ,\quad T^3/\rho \sim \mu^3 m^2\quad ,}\eqno(9)
$$\noindent
where $P$ and $P_{\gamma}$ are the total and radiation pressures.  The first
and second relations give the virial theorem $m/r\sim T/\mu$.
Any given correction $D$ to the ideal gas equation of state
can be included in the homology formulas by replacing $\mu$ with $\mu /(1+D).$
This substitution ignores any $\rho$, $P$, or $T$ dependence in $D$.  Typical
corrections include (Bahcall and Ulrich 1988; Bahcall and Pinsonneault 1992):
the Debye-H\" uckel screening effect, contributing 
$D$ $\simeq -0.014$; photon pressure, contributing $D\simeq$ +0.001; quantum
degeneracy, contributing a negligibly small effect; and a hypothetical
core magnetic field, contributing $D\simeq$ +0.002 for a field strength of
$10^8$ gauss and scaling as the square of the field strength, if the reverse
influence of the thermomechanical structure upon the magnetic field is ignored.

The equations for radiative energy transport and thermal steady state
are, using equation (4b),
$$\eqalign{
\kappa (\ell/m)=4\pi cG (dP_{\gamma}/dP)\quad ,\quad\varepsilon = d\ell/dm\quad 
,}\eqno(10)
$$\noindent
where $\kappa,~\ell,~\varepsilon$ are the Rosseland mean opacity, luminosity, 
specific thermal energy generation at radius $r$, so that 
$$\eqalign{
\ell \sim \mu^4m^3/\kappa\quad .}\eqno(11)
$$\noindent
From equations (9) and (10), we deduce how $\ell$ scales with $\mu,~m,~r:$
$$\eqalign{
\varepsilon\kappa & \sim T^4/P \sim (\mu^2 m)^2\quad.}\eqno(12)
$$\noindent 
The quantity $r^{\nu +3\lambda -s+3n}\sim\varepsilon_o\kappa_o
\mu^{\nu -s-4}m^{\nu +\lambda -s+n-2}$ can be eliminated to obtain
$$\eqalign{
\ell(\mu,m) \sim \varepsilon_o^{-\alpha} \kappa_o^{-\beta} \mu^{\gamma} 
m^{\delta}\quad ,\quad
\ell(\mu,T) \sim \varepsilon_o^{\epsilon} \kappa_o^{\zeta} \mu^{-\eta} T^{\theta}\quad ,}\eqno(13)
$$\noindent
where
$$\eqalign{
 \alpha & = {s-3n\over{\nu +3\lambda -s+3n}}\cr
 \beta & = 1+\alpha\cr
 \gamma & = 4+3n+{(4+3\lambda +3n)(s-3n)\over{\nu +3\lambda -s+3n}}\cr 
 \delta & = 3+2n+{(2+2\lambda +2n)(s-3n)\over{\nu +3\lambda -s+3n}}\quad ,}\eqno(14a)
$$
$$\eqalign{
 \epsilon & = {3+2n\over{2+2\lambda +2n}}\cr
 \zeta & = 1-\epsilon\cr
 \eta & = {(3+2n)(4+3\lambda +3n)\over{2+2\lambda +2n}}-4-3n\cr
 \theta & = {(3+2n)(\nu +3\lambda -s+3n)\over{2+2\lambda +2n}}+s-3n\quad .}\eqno(14b)
$$\noindent
This homology rests on equating the luminosity produced with the luminosity
transported in the steady state~(10).  The exponents in $\ell(\mu,T)$ are 
different from the exponents in $L(\mu, T_{\rm eff})$ obtained when homology is 
applied to an entire star (Cox and Giuli 1968).  We do not consider
the dependence on surface photon temperature, $T_{\rm eff}$, but on
central temperature $T_c$.

For $\lambda=1, ~\nu=4.24$ and the core opacity laws we consider, some numerical
values are given in Table~1.  (The table also contains the exponents for the
B\" ohm-Vitense opacity, with $\nu=4$ for the $ppI$ cycle.)
The exponents are insensitive to the temperature 
exponent $\nu=4.24$ in the luminosity generation law, but are sensitive to the 
opacity law.  For the OPAL opacity, we obtain $\ell\sim m^{4.81}$, in good
agreement with the value $L\sim M^{4.6}$ for the ZAMS Sun in Figure~3.

We are interested in how the temperature varies as function of luminosity
generation, opacity and mean molecular weight, for fixed luminosity $L_\odot.$
$$\eqalign{
T_c\sim (\mu^{\eta}/ \varepsilon_o^{\epsilon} \kappa_o^{\zeta})^{1/\theta}
\quad .}\eqno(15a)
$$\noindent  
For the OPAL opacity function, we obtain the differential relation:
$$\eqalign{
d\ln T_c = (0.215)d\ln\/\mu - (0.133)d\ln\varepsilon_o - (0.0344)d\ln\kappa_o + (0.167)d\ln L_\odot\quad ,}\eqno(15b)
$$\noindent
showing how the central temperature in any compact radiative core must change 
with input parameters.  The central temperature
is most sensitive to the chemically evolved mean molecular weight and to the 
overall luminosity generation $\varepsilon_o,$ and much less sensitive to the 
opacity, $\kappa_o.$
This is expected, since the core structure is determined by mass conservation, 
hydrostatic equilibrium and extended luminosity generation, while the radiative
envelope structure is determined by the radiative transport and the central 
concentration of mass and luminosity.

Because the energy generation is principally proportional to the $ppI$ nuclear
cross section factor, $\varepsilon_o\sim S_{11}$, we obtain $T_c\sim S_{11}^{
-0.134}$, in agreement with Iben (1969; 1991) and Castellani {\it et al.} 
(1993), who, however, 
incorrectly assumed $\rho\sim T^3$.  This explains why in Figures~2, 4, and~6-8 
of Hata and Langacker (1995), $T_c$-parametrization is equivalent to $S_{11}$ 
parametrization, within the uncertainty of either.  The $ppII,$ $ppIII,$ and
$CNO$ chains
contributions to luminosity generation break this simple $T_c$ form, adding
a weak $\rho_c$ dependence.  The density exponents are:
$$\eqalign{
 \rho\sim \varepsilon_o^{-\psi}\kappa_o^{-\sigma}\mu^{\xi}T^{\tau}\quad ,\quad
 \rho\sim \varepsilon_o^{-a}\kappa_o^{-b}\mu^cL_\odot^d\quad ,}\eqno(16)
$$\noindent
with
$$\eqalign{
 \psi & = \sigma = \xi = {1\over{1+\lambda +n}}\cr
 \tau & = {3-\nu +s\over{1+\lambda +n}}\quad ,}\eqno(17a)
$$
$$\eqalign{
    a & = \psi - (\epsilon\tau /\theta )\cr
    b & = \sigma - (\zeta\tau /\theta )\cr
    c & = \xi + (\eta\tau /\theta )\cr
    d & = \tau /\theta\quad .}\eqno(17b)
$$\noindent
The cases of the Kramers and OPAL opacities are presented in Table~2.  The
B\" ohm-Vitense opacity, with $\nu=4,$ is also shown for comparison.

A rotating core would change the hydrostatic equilibrium by adding the 
centrifugal force to that of gravity in the rest frame of solar matter.  If, 
as in the magnetic case, the reverse
influence of the thermomechanical structure on the rotation is neglected,
the correction to the homology is:
$$\eqalign{
T^4/P & \sim \mu^4m^2[1 - \omega ]^3\cr
T^3/\rho & \sim \mu^3m^2[1 - \omega ]^3\quad ,\cr
\ell & = \ell_o(\mu ,T,\kappa_o,\varepsilon_o)[1 - \omega ]^{\chi}\cr
\chi & =  {3(1+\lambda )(3+2n)\over{2(1+\lambda + n)}} - 3\quad ,}\eqno(18)
$$\noindent
where $\ell_o$ is the non-rotating luminosity function, $\omega\equiv$ 
$\Omega^2r^3/Gm,$ and $\Omega$ is the angular rotation frequency.  Near the
center, $\omega\rightarrow$ $3\Omega^2_c/4\pi G\rho_c.$  For typical SSMs,
$\omega_c\approx$ 2$\times$10$^{-7},$ using a reasonable solar core rotation 
rate (Elsworth 1995).  The exponent $\chi$ is given in Table~2 for the 
three opacities.
The rotation correction would be significant only for rotation rates $\sim 
400$ times those in the Sun.  Such a high rotation rate would not only change 
the thermomechanical structure, however, but could also induce sufficient 
chemical mixing to invalidate the homology.

An alternative to this homology is to treat the luminosity $\ell,$ not the
radius $r$ or the cumulative mass $m,$ as the independent variable.  Since
the luminosity is a monotonically increasing function of $r$ in the core, but
not outside, this change of variables is feasible only in the core and 
separates out the luminosity-producing regions.
\bigskip
{\bf 4. Core Homology and Neutrino Fluxes}
\bigskip
After nuclear cross sections are introduced, and the $^3He,~^7Be$ abundances are
assumed to be in steady state, each of the neutrino emissivities, $f_\nu(i)$, 
is a function of ${\bf X},~\rho,~T$.  Homology would then make each neutrino 
flux $\phi (i)\sim f_\nu (i),$ subject to the luminosity constraint~(1).
If there were only one power-law energy generation term in equation~(1), the
core homology would be exact and $\rho_c,$ like all other core variables, 
would be a power of $T_c$ only.  The $Be,$ $B,$ and $CNO$ neutrino production
breakss this homology, so that, besides the principal sensitivity to $T_c,$
the neutrino fluxes have a mild separate dependence on $\rho_c.$ Using 
the luminosity constraint, Gough (1994) has obtained:
$$\eqalign{
\phi(pp) \sim &\ \rho_c^{-0.1}\cdot T_c^{-0.7}\cr
\phi(Be) \sim &\ \rho_c^{0.7}\cdot T_c^{9} \sim \rho_c^{0.57}\cdot\phi(B)^{0.43}\cr
\phi(B) \sim &\ \rho_c^{0.3}\cdot T_c^{21} \sim \rho_c^{-1.33}\cdot\phi(Be)^{2.33}
\quad .}\eqno(19)
$$\noindent
If we approximate $\rho\sim T^3$ in the solar core, we obtain $\phi(i)\sim 
T_C^{\alpha(i)}$, with $\alpha(pp)=-1, ~\alpha(Be)=11,~\alpha(B)=22;$ while 
Castellani {\it et al.} (1993), assuming an 
$n_{\rm eff}=3$ polytropic Sun, obtained $\alpha(pp)=-1.1,~\alpha(Be)=11,
~\alpha(B)=27.$
The small departure from core homology, together with uncertainties in 
the nuclear cross-section factors $S_{34},~S_{17}$, explains the scatter in
diagrams plotting neutrino fluxes against $T_c$ alone (Hata and Langacker
1994; 1995).
\bigskip
{\bf 5. Core Homology and Helioseismology}
\bigskip
Helioseismology, the study of sunquakes, is based on three distinct types of
waves in the solar medium, $p$-modes, $f$-modes, and $g$-modes (Hansen and
Kawaler 1994).  The first
two are acoustic, with pressure contrast as the restoring force, and are
seen in the outer, convective zone, where they have much or most of their 
amplitudes.  Their eigenfrequencies rise with the number of nodes in the
successive modes.

The $g$-mode restoring force is gravity, and these modes have their largest 
amplitude in
the core.  Their eigenfrequencies {\it decrease} with the number of nodes.
The $g$-modes have not yet been firmly detected by optical means, although
they have perhaps been detected through their modulation of the solar
wind (Thomson {\it et al.} 1995).  All modes are labelled by eigennumbers
$n$ and $l,$ with an azimuthal $m$ if rotation is present.  (Otherwise, the
eigenfrequencies are degenerate in $m.$)
The ranges are: $n$ = 1, 2, ..., and $l$ = 0, 1, 2, ....  For large
$n,$ the frequencies of the $g$-modes are given by (Hansen and Kawaler 1994):
$$\eqalign{
\nu_g & = {\sqrt{l(l+1)}\over {2n\pi^2}}\ \Omega_g\quad ,\cr
\Omega_g & =  \int^{R_\odot}_0 dr\ {N(r)\over r}\quad ,\cr
N(r) & = \sqrt{g(r)/\lambda_g(r)}\quad ,}\eqno(20a)
$$\noindent
where $g(r)$ is the local acceleration of gravity and $\lambda_g(r)$ a
local scale height:
$$\eqalign{
{1\over\lambda_g} & = {1\over\lambda_\rho} - {1\over\Gamma_{\rm ad}}\cdot{1\over
\lambda_P} = 
{1\over\lambda_\rho}\Big(1-{\Gamma\over\Gamma_{\rm ad}}\Big)\quad ,}\eqno(20b)
$$\noindent
where $\Gamma_{\rm ad}$ = $(d\ln P/d\ln\rho )_{\rm ad}$ is the adiabatic 
polytropic exponent.

Solar core homology can be applied to the integral $\Omega_g,$ which receives
its main contribution from the core region.  As $r\rightarrow$ 0.
$$\eqalign{
\Omega_g & \simeq 
{4\pi G\over 3}\sqrt{{\rho_c\over P_c}{\Big({1\over\Gamma}-
{1\over\Gamma_{\rm ad}}\Big)_c}}\cdot R_c\cdot\langle\bar\rho(R_c)\rangle
\quad ,}\eqno(21)
$$\noindent
where $R_c$ is the core radius ($R_c=(0.26)R_\odot$) and
$$\eqalign{
\langle\bar\rho(R_c)\rangle & = {1\over R_c}\int^{R_c}_0 
dr\ {3\over{4\pi}}\cdot{m(r)\over r^3}\quad }
\eqno(22)
$$\noindent
is the radially averaged mean density interior to $R_c.$  Note $\Gamma_{\rm ad}
\geq \Gamma$ implies convective stability of the core.

Because, outside the immediate central region $(x<0.049),$ the mass $m(r)$
rises more slowly than $r^3,$
the integral emphasizes the central core as the dominant 
``yolk in the egg'' mass concentration that controls the $g$-mode oscillations. 
A simple estimate is $\langle\bar\rho(R_c)\rangle\approx\rho_c,$ but 
a better estimate results from
applying core homology via the $n_{\rm eff}=\infty$ ``isothermal'' polytropic
solution~(Chandrasekhar 1939; Kippenhahn and Weigert 1990).  The important 
length scale here is $\sqrt{P_c/4\pi G\rho^2_c}
=0.049R_\odot .$  Using the power series and asymptotic properties of
the solution, one obtains:
$$\eqalign{
\langle\bar\rho (R_c)\rangle & = {3\rho_c\over z_c}\int^{z_c}_0dz
\ {dw(z)/dz\over z}\cr
                      & \simeq 0.56\rho_c\quad ,}\eqno(23)
$$\noindent
where $z_c=0.26/0.049=5.3$ and $w(z)$ is the dimensionless 
gravitational potential.  This estimate is smaller and more accurate than
$\rho_c$ as it covers the entire core, whose average density is lower
than its central density.

As helioseismological observations are so far in good agreement with SSM 
predictions,
we conclude that the homology presented in this paper is, for practical
purposes, a complete parametrization for any ``reasonable'' changes to the
minimal SSM. 
\vfill\eject
{\bf 6. Conclusion}
\bigskip
Assuming mechanical and thermal stasis and neglecting chemical evolution, the
homology makes the
$(T_c,\rho_c)$ parametrization a general framework characterizing the
inner core.  Using this approach,
we have disposed of three misconceptions: (1) that the luminosity-generating 
core of the
Sun is polytropic; (2) that the polytropic relation $\rho\sim T^3$ is essential
to understanding the Sun's core; and (3) that homology is inapplicable to stars 
on the middle of the main sequence.  While the $T$-gradient depends
on the opacity, $T_c$ depends mainly on the mean molecular weight
$\mu$ because of the homology, assuming a quiescent non-convective and 
unmixed core.  The properties of the core depend only on
one surface boundary condition, the total luminosity $L_\odot$, assumed to be
in steady state with the core's luminosity.  The surface temperature $T_{\rm
eff}$ is then irrelevant.

The solar core is almost homologous because its luminosity generation is 
dominated by the $ppI$ cycle and, over its narrow range of temperature and 
density, the opacity and luminosity generation can be approximated by power 
laws.  The luminosity of solar models based on purely radiative transfer scales
by the nuclear cross-section factor $S_{11}$, or equivalently, by $T_c\sim 
S_{11}^{-0.14}$.  This quasi-homology justifies the $T_c$-parametrization as
reasonable for estimating astrophysical uncertainties in any SSM and for 
extrapolating from any SSM to even extreme non-standard ``cool Sun'' or ``hot 
Sun'' models.  In particular, analyses such as those of Hata and Langacker 
(1994; 1995)
and of Bludman, Hata, Kennedy, and Langacker (1993), using
the $T_c$-parametrization and nuclear cross section uncertainties alone, can
arrive at theoretical neutrino flux and detection rate uncertainties and their 
correlations, agreeing with those Bahcall and Ulrich obtained from 1000
different Monte Carlo SSM simulations (Bahcall and Ulrich 1988).

Because of the central concentration of mass and luminosity generation, the 
Sun's outer radiative zone is nearly an $n_{\rm eff}=4.3$ partial polytrope, 
with 
exponential pressure, density, and temperature profiles.  The density scale 
height $\lambda_\rho = 0.095 R_\odot$ is the single solar parameter entering 
into the MSW adiabaticity parameter that determines any small-angle MSW 
oscillations.
\bigskip
{\bf Acknowledgements}
\bigskip
We thank David Dearborn of Lawrence Livermore National Laboratory for
providing his SSM results used in Figures~1 and 2 and Naoya Hata of Ohio
State University for preparing these figures.
This research was supported by the Department of Energy under Grant Nos.
DE-FG05-86-ER40272 (Florida) and DE-AC02-76-ERO-3071 (Penn) and by the
National Science Foundation under Grant No. PHY89-04035 (U. California,
Santa Barbara).  We thank the Institute for Nuclear Theory (ITP),
University of Washington at Seattle and the Aspen Center for Physics for their
hospitality.  D.C.K. also thanks the Institute for Theoretical
Physics (UCSB).  
\bigskip
{\bf Appendix: The Luminosity Constraint}
\bigskip
The constraint of the total photon luminosity $L_\odot$ upon the neutrino
fluxes is almost independent of the specific SSM, resulting for the most
part from the microscopic properties of the nuclear fusion reactions 
(Schwarzschild 1958; Cox and Giuli 1968; Bahcall and Ulrich 1988; 
Turck-Chi\` eze 1988).  The branching ratios assumed are mildly
SSM-dependent and are taken from the results of Bahcall and Pinsonneault 
(1992), with helium diffusion.  This luminosity constraint on the neutrino
fluxes should be distinguished from that used to determine the helium
abundance $Y.$

The Sun shines by two nuclear reaction chains, the dominant $pp$ and the
minor $CNO$.  The $pp$ chain itself consists of three subchains, $ppI$, $ppII$,
and $ppIII$, each terminating in ${}^4He$ or $\alpha$ production in a 
different way.
$$\eqalign{
\matrix{   & & {}^3He{}^3He & \rightarrow & \alpha & & & & & \cr
           & & \uparrow & & & & & & & \cr
    pp/pep & \rightarrow & & & & & & & & \cr
           & & \downarrow & & & & & & & \cr
           & & {}^3He{}^4He & \rightarrow & {}^7Be & \rightarrow & {}^7Li &
\rightarrow & 2\alpha\cr
           & & & & \downarrow & & & & & \cr
           & & & & {}^8B & \rightarrow & 2\alpha & & & \cr}}
$$\noindent
For each reaction, the energy released, $Q,$ is partitioned between
neutrino energy $Q_\nu$ and photon luminosity 
$Q_\gamma$, so that $Q$ = $Q_\gamma$ + $Q_\nu$.  Reactions without neutrinos
have $Q_\nu$ = 0.  

The luminosity constraint arises from the proportionality of the neutrinos
fluxes to the nuclear reaction rates.  Each reaction contributes its $Q_\gamma$
value to the photon luminosity.  Since there are {\it two} neutrinos emitted for
each reaction chain, the luminosity is related to the neutrino fluxes by
$$\eqalign{
L_\odot = \Big(\sum_i Q_\gamma (i)\phi(i)\Big)\cdot (4\pi R^2_\odot/2)\quad ,}
\eqno(A.1)
$$\noindent
summed over all nuclear reactions $i$.  Normalizing to the $ppI$ $Q_\gamma$
value and to the neutrino fluxes measured at the Earth's orbit,
$$\eqalign{
{\sum_i Q_\gamma (i)\phi (i)\over{Q_\gamma (ppI)}} =
{2\over 4\pi R^2_\odot}\cdot{L_\odot\over{Q_\gamma (ppI)}}\cdot\Big({R_\odot
  \over r_\oplus}\Big)^2\quad ,}\eqno(A.2)
$$\noindent
where $r_\oplus$ = 149.6$\times$10$^6$ km is the Earth's average orbital radius.

The $Q$, $Q_\nu$, and $Q_\gamma$ values for all four reaction chains are listed
in Table~3 (Fowler 1967; Bahcall and Ulrich 1988; Turck-Chi\` eze {\it et al.} 
1988).  The $pp$ chain is initiated by either the $pp$ or $pep$ reactions,
in the ratio 99.6\%~:~0.4\%; the $ppII$ chain emits neutrinos at two discrete 
energies, in the ratio 89.7\%~:~10.3\%.  The quoted energies are weighted 
averages.  The ratios of the $Q_\gamma$ values can then be computed to obtain
the constraint~(1).

Also necessary are the ratios of the different neutrino fluxes in order to
obtain equation~(2,3).  These fluxes and the percentages above are obtained
from a specific SSM, but the coefficients in the constraints~(1-3) are
relatively insensitive to theoretical variations that do not depart radically 
from conventional SSMs.
The ratios of the reaction subchain termination rates are
related to the fluxes by:
$$\eqalign{
 {{\rm term}(ppII)\over{\rm term}(ppI)} & = {\phi(Be)\over{[\phi(pp)/2] - 
    \phi(Be)}}\quad ,\cr
 {{\rm term}(ppIII)\over{\rm term}(ppII)} & = {\phi(B)\over{\phi(Be)}}\quad .}
\eqno(A.3)
$$\noindent
\vfill\eject

\centerline{TABLE 1}
\bigskip
\settabs 8 \columns
\centerline{\it Luminosity Power Laws$^{\rm a}$}
\medskip
\hrule
\smallskip
\+ Opacity & $n$ & $s$ & $\delta$ & $\epsilon$ & $\zeta$ & $\eta$ & $\theta$\cr
\smallskip
\hrule
\smallskip
\+ Kramers & 1 & 3.5 & 5.44 & 0.833 & 0.167 & 1.33 & 6.12\cr
\medskip
\+ BV & 0.5 & 2.5 & 4.83 & 0.8 & 0.2 & 1.3 & 5.8\cr
\medskip
\+ OPAL	& 0.43 & 2.47 & 4.81 & 0.794 & 0.206 & 1.29 & 5.99\cr
\smallskip\smallskip
\hrule
\medskip
\midinsert\indent{\it $^{\rm a}$Exponents in the luminosity power-laws (15) 
for three Rosseland mean opacities of the form (7).}\endinsert
\vskip 1cm
\settabs 8 \columns
\centerline{TABLE 2}
\bigskip
\centerline{\it Density Power Laws$^{\rm a}$}
\medskip
\hrule
\smallskip
\+ Opacity & $\psi$ & $\tau$ & $a$ & $b$ & $c$ & $d$ & $\chi$\cr
\smallskip
\hrule
\smallskip
\+ Kramers & 0.333 & 0.753 & 0.231 & 0.313 & 0.497 & 0.123 & 2\cr
\medskip
\+ BV & 0.4 & 0.6 & 0.317 & 0.372 & 0.534 & 0.103 & 1.8\cr
\medskip
\+ OPAL & 0.412 & 0.507 & 0.345 & 0.395 & 0.521 & 0.0847 & 1.77\cr
\smallskip\smallskip
\hrule
\medskip
\midinsert\indent{\it $^{\rm a}$Exponents in the density power-laws (16) for 
three
Rosseland mean opacities of the form (7).  See text for last entry $(\chi ).$}
\endinsert
\vskip 1cm
\centerline{TABLE 3}
\bigskip
\settabs 4 \columns
\centerline{\it $pp$ Reaction Q Values$^{\rm a}$}
\smallskip
\hrule
\smallskip
\+ Chain & $Q$ & $Q_\nu$ & $Q_\gamma$\cr
\smallskip
\hrule
\smallskip
\+ $\ \ ppI(pp)$ & 26.732 & 0.540 & 26.192\cr
\smallskip
\+ $ppII(Be)$ & 26.682 & 1.083 & 25.599\cr
\smallskip
\+ $ppIII(B)$ & 26.639 & 6.980 & 19.659\cr
\smallskip
\hrule
\medskip
\midinsert
\indent{\it $^{\rm a}$All energies in {\rm MeV.}  $Q_\nu$ is spectrum-averaged 
for each reaction chain.}\endinsert
\vfill\eject

\noindent{\bf References}
\bigskip
\item{} Bahcall, J.~N. and Pinsonneault, M.~H. 1992, {\it Rev. Mod. Phys.} 
{\bf 64}, 885.

\item{} Bahcall, J.~N. and Ulrich, R.~K. 1988, {\it Rev. Mod. Phys.} {\bf 60}, 
297.

\item{} Bludman, S.~A. {\it et al.} 1993, {\it Phys. Rev.} {\bf D47}, 2220.

\item{} B\" ohm-Vitense, E. 1992, {\it Introduction to Stellar Astrophysics,} 
vol.~3 (Cambridge: Cambridge University Press).

\item{} Castellani, V. {\it et al.} 1993, {\it Phys. Lett.} {\bf B303}, 68 and 
{\it Astron. Astrophys.} {\bf 271}, 601.

\item{} Chandrasekhar, S. 1939, {\it An Introduction to the Study of Stellar 
Structure} (Chicago: University of Chicago).

\item{} Cox, A.~N. and Guili, R.~T. 1968, {\it Principles of Stellar Structure,}
vol.~II (New York: Gordon and Breach).

\item{} Dearborn, D. 1994, private communication.

\item{} Degli'Innocenti, S. 1994, private communication and in {\it Proc. Solar
Modeling Workshop} (Seattle: U. Washington/Institute for Nuclear Theory).

\item{} Elsworth, Y. {\it et al.} 1995, {\it Nature} {\bf 376}, 669.

\item{} Fowler, W.~A. 1967, {\it Nuclear Astrophysics} (Philadelphia: American 
Philosophical Society).

\item{} Gough, D.~O. 1994, {\it Phil. Trans. R. Soc. London} {\bf A346}, 37.

\item{} Hansen, C.~J. and Kawaler, S.~D. 1994, {\it Stellar Interiors: Physical 
Principles, Structure, and Evolution} (Berlin: Springer-Verlag).

\item{} Hata, N. 1994, in {\it Proc. Solar Modeling Workshop} (Seattle: 
U. Washington/Institute for Nuclear Theory).

\item{} Hata, N. {\it et al.} 1994, {\it Phys. Rev.} {\bf D49}, 3622.

\item{} Hata, N. and Langacker, P. 1994, {\it Phys. Rev.} {\bf D50}, 632.

\item{} Hata, N. and Langacker, P. 1995, {\it Phys. Rev.} {\bf D52}, 420.

\item{} Iben, I., Jr. 1969, {\it Ann. Phys.} (NY) {\bf 54}, 164.

\item{} Iben, I., Jr. 1991, {\it Ap. J. Suppl.} {\bf 76}, 55.

\item{} Iglesias, C.~A. and Rogers, F.~J. 1991, {\it Ap. J.} {\bf 371}, 408.

\item{} Kippenhahn, R. and Weigert, A. 1990, {\it Stellar Structure and 
Evolution} (Berlin: Springer-Verlag).

\item{} Rogers, F.~J. 1992, {\it Ap. J. Suppl.} {\bf 79}, 507.

\item{} Schwarzschild, M. 1958, {\it Structure and Evolution of the Stars} 
(Princeton: Princeton University Press).

\item{} Thomson, D.~J. {\it et al.} 1995, {\it Nature} {\bf 376}, 139.

\item{} Turck-Chi\` eze, S. {\it et al.} 1988, {\it Ap. J.} {\bf 335}, 415.

\vfill\eject

{\bf Figure Captions}
\bigskip
\item{Fig.~1.} The stiffness coefficient $d\ln P/d\ln\rho$ across the Sun's
profile.  In the convective zone, $\Gamma=5/3;$ in the outer radiative zone,
$\Gamma\approx 1.23;$ but over the core, $\Gamma\sim 1$ and varies.

\item{Fig.~2.} The temperature gradient $\nabla = d\ln T/d\ln P$ and 
$1-\Gamma^{-1}=d\ln (P/\rho )/d\ln P.$  Where the chemical composition is
homogeneous, $\nabla =1-\Gamma^{-1}=0.4$ in the convective zone and $\approx
0.18$ in the outer radiative zone.  But over the core, $P/(\Re\rho T)=
(1+D)/\mu$ is varying.

\item{Fig.~3.} The lines shows calculated $L$--$M$ and $R$--$M$ relations for a
large range of zero-age Main Sequence stars.  The dots and triangles show best
measurements of slected Main Sequence components of detached and visual
binary components, respectively.  (From Kippenhahn and Weigert (1990), by
permission.)

\vfill\eject
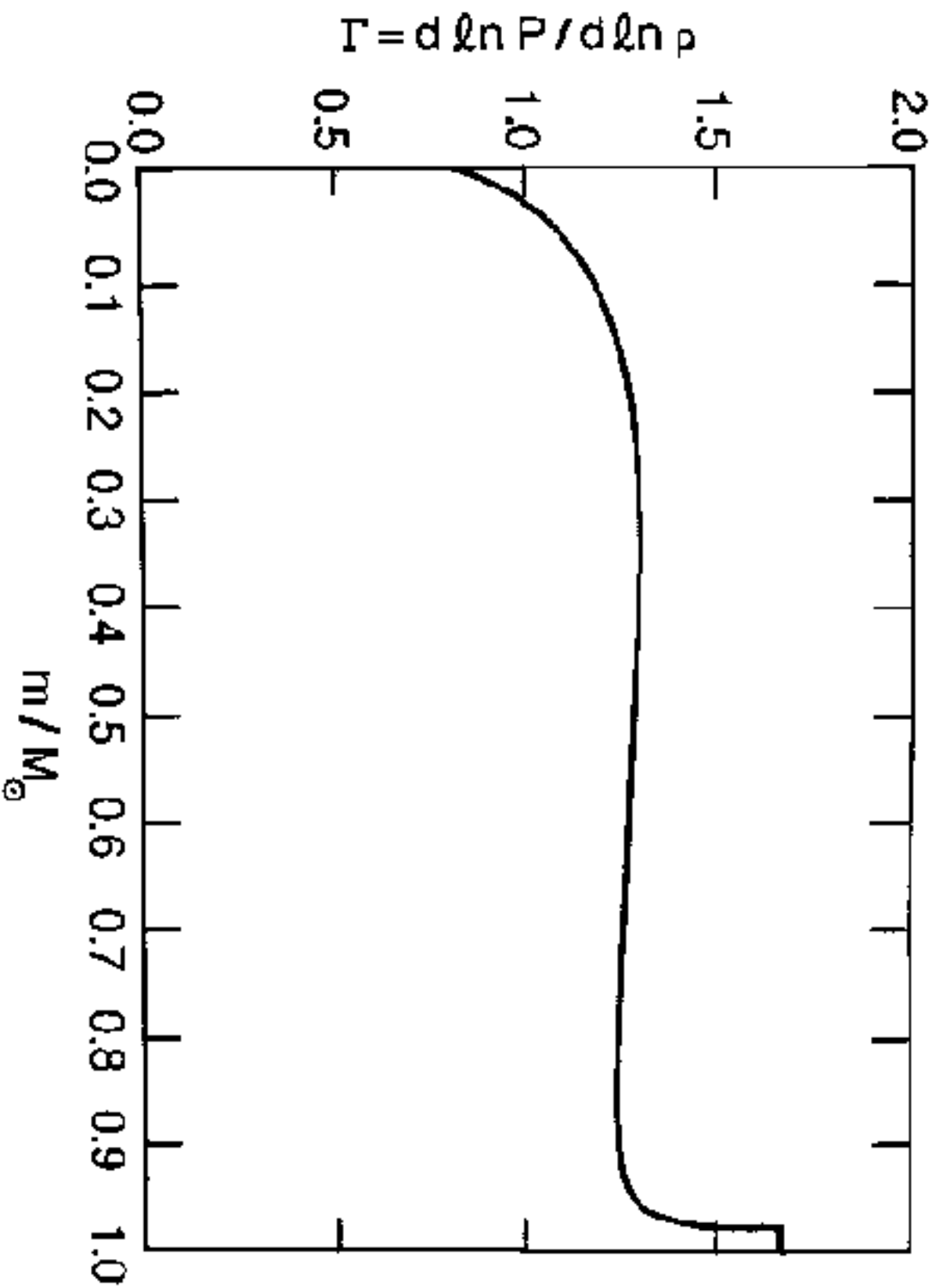
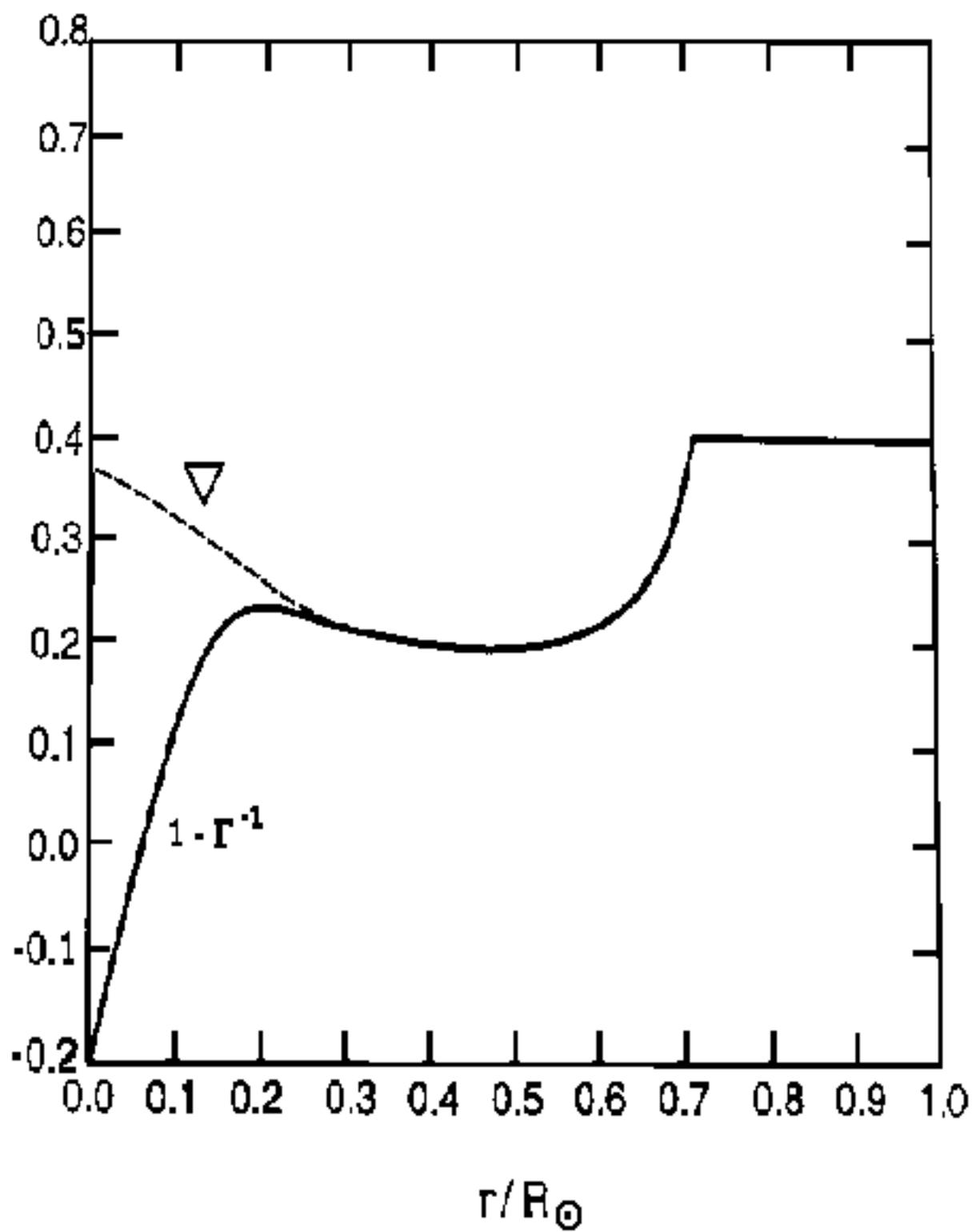
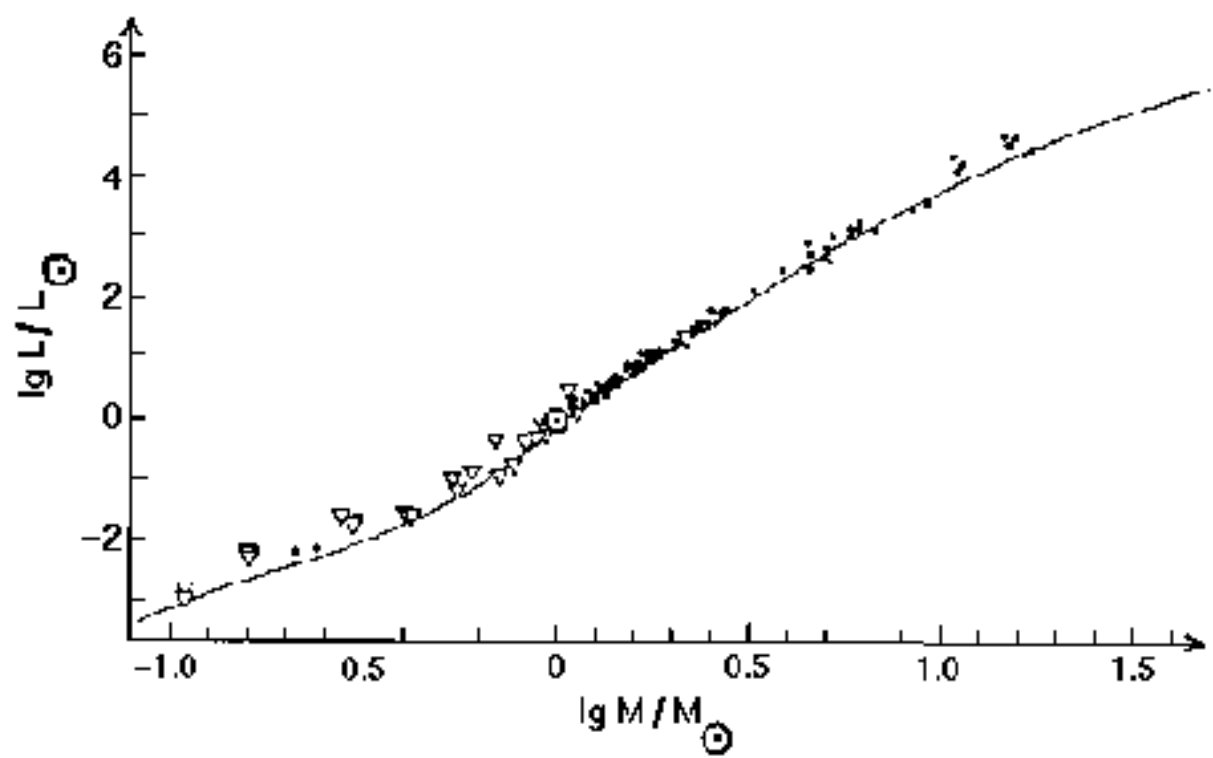
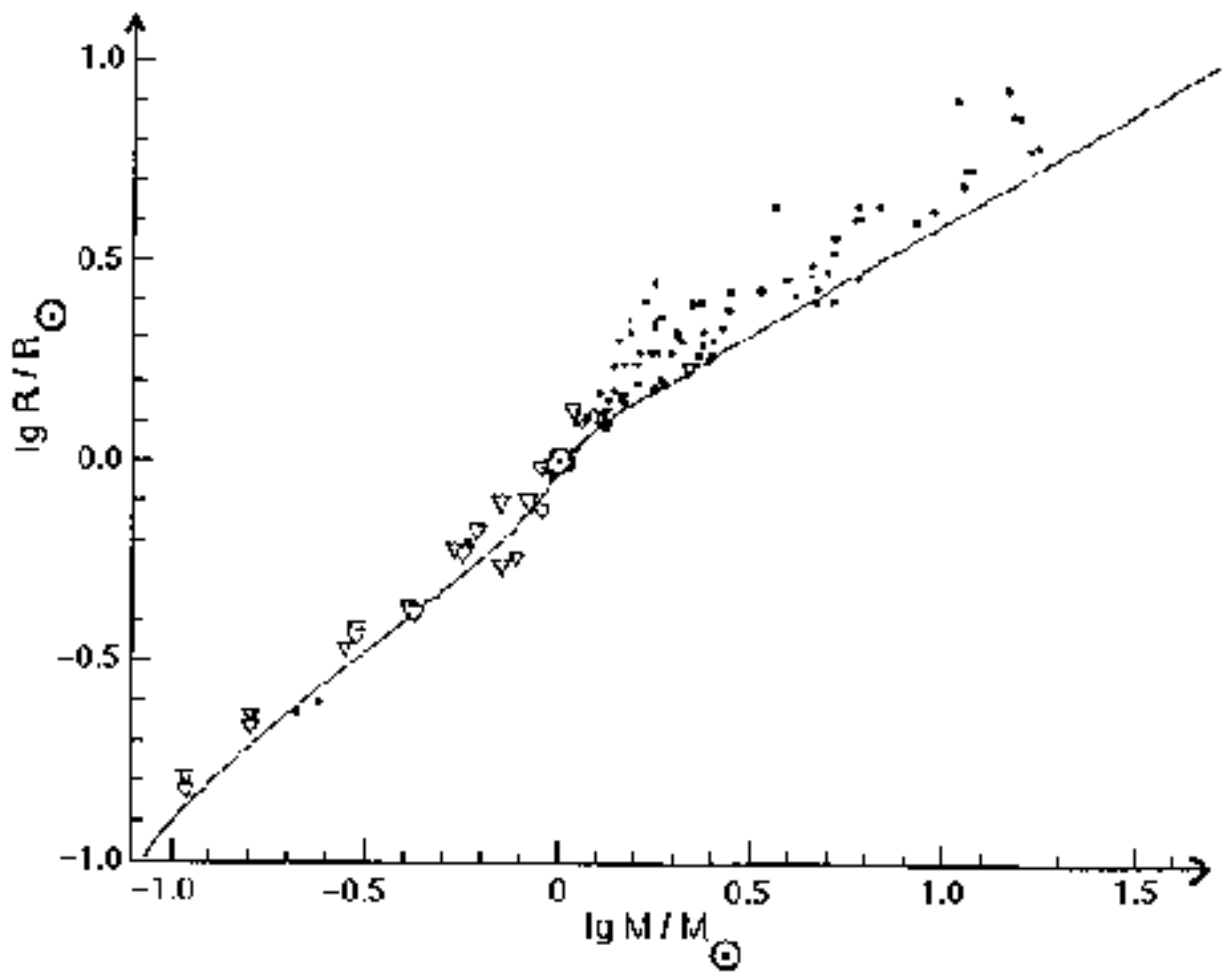
\end